\documentclass[reprint,preprintnumbers,nofootinbib,amsmath,amssymb,aps,pra]{revtex4-2}
\pdfoutput=1
\usepackage{graphicx,hyperref,braket,orcidlink, algorithm2e}
\usepackage{lineno}
\graphicspath{{figs/}}
\hypersetup{
	bookmarks=true,
	unicode=true,
	pdftoolbar=true,
	pdfmenubar=true,
	pdffitwindow=false,
	pdfstartview={FitH},
	pdftitle={Quantum-probabilistic Hamiltonian learning for generative modelling and anomaly detection},
	pdfauthor={J.Y.~Araz, M.~Spannowsky}, 
	pdfsubject={Quantum-probabilistic Hamiltonian learning for generative modelling and anomaly detection},
	pdfcreator={J.Y.~Araz, M.~Spannowsky},
	pdfproducer={}, 
	pdfkeywords={Quantum Machine Learning},
	pdfnewwindow=true,
	colorlinks=true,
	linkcolor=blue,
	citecolor=magenta,
	filecolor=magenta,
	urlcolor=cyan
}

\newcommand{\tr}[1]{\mathrm{Tr}\left[#1\right]} % Trace
\newcommand{\Unitary}{\hat{U}(\phi)} % U(phi)
\newcommand{\Unitaryd}{\hat{U}^{\dagger}(\phi)} % U^\dagger(phi)
\newcommand{\rthp}{\hat{\rho}_{\theta, \phi}} % \rho_\theta
\newcommand{\sigD}{\hat{\sigma}_\mathcal{D}} 
\newcommand{\sigd}{\hat{\sigma}_d} 

\newcommand{\Kth}{\hat{\mathcal{K}}_\theta}
\newcommand{\expK}{\langle\hat{\mathcal{K}}\rangle_{\theta,\phi}}
\newcommand{\state}[1]{|#1\rangle} % Trace
\newcommand{\stateinv}[1]{\langle#1|} % Trace
\begin{document}

%\title{Hybrid quantum--probabilistic learning for High Energy Physics}
%\title{Generative modelling \& anomaly detection\\ with hybrid quantum--probabilistic learning}
\title{Quantum--probabilistic Hamiltonian learning\\ for generative modelling \& anomaly detection}

\author{Jack Y. Araz\orcidlink{0000-0001-8721-8042}}
\email{jack.araz@durham.ac.uk}
\affiliation{Institute for Particle Physics Phenomenology, Durham University, South Road, Durham DH1 3LE, United Kingdom}

\author{Michael Spannowsky\orcidlink{0000-0002-8362-0576}}
\email{michael.spannowsky@durham.ac.uk}
\affiliation{Institute for Particle Physics Phenomenology, Durham University, South Road, Durham DH1 3LE, United Kingdom}

%\date{\today}

\preprint{IPPP/22/77}

% \linenumbers
\begin{abstract}
The Hamiltonian of an isolated quantum mechanical system determines its dynamics and physical behaviour. This study investigates the possibility of learning and utilising a system's Hamiltonian and its variational thermal state estimation for data analysis techniques. For this purpose, we employ the method of Quantum Hamiltonian-Based Models for the generative modelling of simulated Large Hadron Collider data and demonstrate the representability of such data as a mixed state. In a further step, we use the learned Hamiltonian for anomaly detection, showing that different sample types can form distinct dynamical behaviours once treated as a quantum many-body system. We exploit these characteristics to quantify the difference between sample types. Our findings show that the methodologies designed for field theory computations can be utilised in machine learning applications to employ theoretical approaches in data analysis techniques.
\end{abstract}

\maketitle

%%%%%%%%%%%%%%%%%%%%%%%%%%%%%%%%%%%%%%%%%%%%%%%%%%%%%%%%%%%
%\section*{Introduction}\label{sec:intro}
%%%%%%%%%%%%%%%%%%%%%%%%%%%%%%%%%%%%%%%%%%%%%%%%%%%%%%%%%%%

The Hamiltonian plays a crucial role in our theoretical understanding of a physical system. The dynamics of an isolated quantum system are governed by an effective Hamiltonian which indicates the interaction of the system's constituents. In many, arguably simple, cases, it is possible to determine the effective Hamiltonian through a set of theoretical considerations, such as observing its interactions and using the underlying symmetries of the system. Often, however, it is challenging to derive the algebraic form of a Hamiltonian from theoretical considerations only. Hence, several Hamiltonian learning methods have been proposed by employing thermal or eigenstates~\cite{PhysRevLett.122.020504, Bairey_2020, Anshu:2021aa, Haah:2021nzn, Pizorn_2014, Qi2019determininglocal}, short-time evolutions~\cite{https://doi.org/10.48550/arxiv.2108.08824, PRXQuantum.2.010322, Franca:2022jwx, Gu:2022exn}, and data-driven approaches~\cite{Wilde:2022uly}. With recent technological developments, it has become possible to simulate the effective Hamiltonian that governs a quantum many-body system in an actual quantum device. Widely used methods like Variational Quantum Eigensolver (VQE)~\cite{Peruzzo:2014aa, McClean_2016} and its generalisation Variational Quantum Thermaliser (VQT)~\cite{Verdon:2019} have become the most promising algorithms for noisy-intermediate scale quantum (NISQ) devices~\cite{Preskill2018quantumcomputingin}.

There is a substantial yet often underappreciated similarity between the computational methods used for data analysis, {\it e.g.} in quantum machine learning and the theoretical description of quantum many-body systems. In both scenarios, one optimises the variational parameters of a given ansatz over an objective function. For the former, this is naturally the expectation value of a Hamiltonian, and for the latter, it is a loss function chosen for the nature of the problem. In Machine Learning (ML) applications, one usually chooses one or multi-qubit measurement with a Pauli operator and optimises the probability of the expectation value of this operator on a set of qubits. However, this operator is not necessarily optimal for the optimisation process, where it is only a subset of possible combinations of different operators. In this study, we will investigate the possibility of learning an optimal effective operator for the optimisation process and the implications of this operator for the application.

In generative modelling, the aim is to learn the joint probability distribution between the target and the observed data, which enables the model to generate new data resembling the observed data. This requires representing the probability distribution of the data within a quantum device. The mixed states are an ideal surrogate for such representation since they form probabilistic mixtures of pure states. Additionally, mixed states attain the properties of both quantum and classical correlations, enhancing the representability of a given probability distribution. The likeness of a given probability distribution can be captured within a Parametrised Quantum Circuit (PQC) as a thermal state of a modular Hamiltonian. Quantum Hamiltonian-Based Models (QHBM)~\cite{Verdon:2019} have been proposed as generative models which split the learning process into two distinct parts. The first part is responsible for learning a modular Hamiltonian with the aid of a classical neural network for capturing the classical correlations within the data. The second portion consists of a PQC that constructs the learned Hamiltonian's thermal state. The aim is to approximate the probability distribution of the data by optimising the learned thermal state with respect to the mixed state based on the data.

Motivated by the close methodical relation between data analysis and the simulation of field theories, {\it i.e.} the use of optimisation methods on parametrised circuits, we propose to learn a Hamiltonian from data, according to the QHBM approach. Thus, we will demonstrate an end-to-end hybrid quantum--probabilistic optimisation procedure to simultaneously learn the probability distribution and modular Hamiltonian for the data. 
We then use the learned probability distribution to generate new data and, as a case study, apply it to top-quark production at the Large Hadron Collider (LHC). Furthermore, we will use the learned Hamiltonian for anomaly detection, as we will show that both the expectation value of the learned Hamiltonian and the Hamiltonian-based time evolution sequence discriminate between signal and background data samples. 

Our findings in this study show that the optimisation methods developed to simulate quantum many-body systems are easily transferable to data-analysis applications and can be used to integrate the theoretical foundations of quantum mechanics into ML techniques and vice versa. To our knowledge, this approach has not been proposed or investigated for data-analysis applications\footnote{In previous studies, the generative modelling has been used in the context of the Quantum Generative Adversarial Networks (Q-GAN)~\cite{Zoufal:2019aa, Assouel:2022aa, BravoPrieto2022stylebasedquantum, Chang:2022dxc}, and anomaly detection has been presented via PQC~\cite{Alvi:2022fkk} and quantum variational autoencoders~\cite{Ngairangbam:2021yma}. See ref.~\cite{Blance:2021aa} for an example of quantum machine learning for particle physics.}.

Previous studies on Hamiltonian learning techniques are mainly based on learning the structure of a quantum state, e.g. with simulated data from a known Hamiltonian with certain noise~\cite{Wilde:2022uly}. The initial proposal of QHBM~\cite{Verdon:2019} is designed to provide a hybrid learning algorithm. However, due to the complexity of the quantum systems, it is challenging for a classical computer to generate efficient enough samples. To alleviate that problem, various solutions have been proposed, such as approximated free energy techniques~\cite{bassman2021computing} and replacing classical sampling of the quantum data with a quantum circuit that represents the sample itself~\cite{Selisko:2022wlc}. Simulating thermal states are especially challenging due to their circuit depth requirements which can be remedied via noise-assisted thermal state preparation~\cite{Foldager:2021qyk}. Additionally, various density matrix simulation algorithms have been proposed for error mitigation~\cite{Cerezo:2020zcm}, and algorithms that are enhanced with classical post-processing techniques to simulate non-trivial Hamiltonian~\cite{Zhang:2021abt}. Despite the plethora of applications for quantum simulation, such techniques have not been employed in data analysis methodology, where the closest application was building a covariance matrix for principal component analysis through density matrices~\cite{Gordon:2022obe}. Our implementation goes beyond quantum simulations and discusses the usage of such techniques for data analysis by representing generic data as a mixed quantum state of a learned operator and, in doing so, generating an abstract representation of the entire dataset as a Hamiltonian.

%%%%%%%%%%%%%%%%%%%%%%%%%%%%%%%%%%%%%%%%%%%%%%%%%%%%%%%%%%%

This study has been structured as follows; in section~\ref{sec:qhbm} we outline the methodology that is adapted. Section~\ref{sec:results} introduces the dataset and preprocessing scheme which is followed by generative modelling exercise in section~\ref{sec:generative_modeling} and anomaly detection in section~\ref{sec:anomaly}. Finally, we offer conclusions in section~\ref{sec:conclusion}.

%%%%%%%%%%%%%%%%%%%%%%%%%%%%%%%%%%%%%%%%%%%%%%%%%%%%%%%%%%%
\subsection{Quantum Hamiltonian-Based Models}\label{sec:qhbm}
%%%%%%%%%%%%%%%%%%%%%%%%%%%%%%%%%%%%%%%%%%%%%%%%%%%%%%%%%%%

This section will be subdivided into three distinct subsections. Firstly, we will delve into the intricacies of constructing the quantum variational ansatz, emphasizing the incorporation of data embedding. Secondly, we will elucidate the construction of the Hamiltonian, providing rationale for the specific approach chosen. Lastly, we will delineate the formulation of the objective function, drawing parallels to classical generative modeling for a clearer understanding.

\subsubsection{Quantum variational ansatz \& data representation}

The objective of generative modeling is to encapsulate the entire feature space within a single probability distribution function, enabling efficient sampling to replicate the underlying distribution. Quantum mechanics provides a natural framework for this representation through the concept of mixed states. In quantum theory, a mixed state is a composite entity formed by combining pure states or other mixed states, serving as a faithful representation of the probability distribution encompassing its constituent states. A mixed state can be represented as
\begin{eqnarray}
    \sigma=\sum_ip_i\state{s_i}\stateinv{s_i}\ ,
\end{eqnarray}
where $p_i$ is the probability of observing the state $\state{s_i}$ in the mixed state $\sigma$. Hence $\sum_i p_i = 1$. If we aim to acquire a density matrix representation of the feature space for data regeneration, a challenge arises due to the inherent nature of quantum circuits as pure state simulators. Embedding a mixed state directly onto a quantum circuit is not straightforward. To circumvent this challenge, we can reinterpret the data samples as a probability distribution. Each feature within a data point possesses an associated occurrence probability, which can be effectively encoded onto the quantum circuit using binary values (0's and 1's) derived from sampling a Bernoulli distribution. By generating a sufficient number of such samples, the mean of the sample set corresponds to the occurrence probability of the feature. Through the learning of these samples, it becomes possible to reconstruct the correlation relationships between the features within the quantum circuit.

By this method a single data point will be represented as a collection of pure states which sampled from a Bernoulli distribution,
\begin{eqnarray}
    \state{p_n}^i_d\equiv\state{p_1,\cdots,p_n}^i_d =& {\rm Bernoulli}(p_1,\cdots,p_n)\ ,\label{eq:bernoulli_sample}
\end{eqnarray}
where the state for $d$-th data point with $n$ individual feature probabilities given as $p_i$, has been represented as the sample $i$, drawn from Bernoulli distribution. This state can be embedded on a quantum circuit by applying Pauli-$X$ gates where Bernoulli distribution results in 1. The combination of the entire data set in terms of sampled states can be represented as, 
\begin{eqnarray}
    \state{\mathcal{D}} = \sum_{d\in\mathcal{D}}\frac{\alpha_d}{N}\sum_i^N \state{p_n}^i_d\ , \label{eq:sigma_n}
\end{eqnarray}
where $\alpha_d$ is the weight of each data point within the data set $\mathcal{D}$. Here $N$ represents the number of samples drawn from the Bernoulli distribution for each data point. Finally, the mixed state of the entire data set can be represented as $\sigD = \state{\mathcal{D}}\stateinv{\mathcal{D}}$ with appropriate normalization. Notice that, now a data point does not correspond to a single circuit measurement but of a stack of circuits with different binary inputs from a single Bernoulli distribution. Furthermore, one can learn the correlation structure of the by means of a variational ansatz, $\hat{U}(\phi)$, where $\phi$ represents the trainable parameters of the ansatz.

\subsubsection{Building the Hamiltonian}

To facilitate the optimization process, it is imperative to establish a well-defined measurement protocol. This protocol must employ an operator or Hamiltonian capable of accurately encapsulating the entropic probability distribution associated with the targeted mixed state for acquisition. While one option involves the application of a pre-existing Hamiltonian, such as the Ising model, it is essential to acknowledge that this approach inherently assumes specific correlation structures among the features. Alternatively, a more ambitious endeavor entails simultaneously acquiring both the complete Hamiltonian representation and the density matrix characterizing the underlying data through an optimization process.

Given the resource-intensive nature of learning a Hamiltonian, we have opted for the utilization of a classical neural network, which offers a more flexible structural framework. However, it is worth noting that while any neural network ansatz can be employed for this purpose, the estimation of the partition function may pose computational challenges that exceed the allocated resources. The necessity of a partition function will become evident in the subsequent subsection.

Energy-Based Models (EBMs)~\cite{https://doi.org/10.48550/arxiv.1903.08689} present an ideal choice for our task, as they inherently encompass the partition function. An EBM establishes a mapping between a state configuration and a scalar energy measure, denoted as $E_\theta(v):= v\in\mathcal{V}~\to~\mathbb{R}$, where $v\in\mathcal{V}$ represents a spin configuration within the set of all possible configurations. EBMs are designed to determine the optimal energy by minimizing the marginal probability distribution of the states in $\mathcal{V}$, given by
\begin{eqnarray}
    p(v) = \frac{1}{\mathcal{Z}_\theta}\ e^{-E_\theta (v)}\quad ,\quad \mathcal{Z}_\theta=\sum_{v\in\mathcal{V}} e^{-E_\theta (v)} \ , \nonumber
\end{eqnarray}
Computation of the energy and partition function for all possible state configurations is generally a formidable task. To address this, we employ Monte Carlo (MC) algorithms to sample states, accepting the most probable ones based on the partition-free acceptance rate defined as $\min(p(v_{n+1}) / p(v_n),\ 1)$ for a randomly initialized state $v_{n+1}$ and a previously chosen random state $v_n$. By doing so, the MC algorithm generates an ensemble of $|v_n\rangle$, i.e. Gibbs state. This algorithm can be visualised by the following pseudo-code

\begin{algorithm}[H]
 initial state: $ v_0 $\;
 \While{$n<N^{\rm MC}$}{
  propose a state: $v_n$\;
  \eIf{$\min(p(v_{n}) / p(v_{n-1}),\ 1) < $ Random number}{
   add to the Gibbs state\;
   increment $n$\;
   }{
   continue\;
  }
 }
 \caption{\it The MC algorithm to form the modular Hamiltonian.}\label{code:mc}
\end{algorithm}

With a sufficient number of MC samples, $N^{\mathrm{MC}}$, we define a modular Hamiltonian as follows:
\begin{eqnarray}
    \hat{\mathcal{K}}_\theta = \sum_{n=1}^{N^{\mathrm{MC}}} E_\theta(v_n) \state{v_n}\stateinv{v_n} \ , \label{eq:hamiltonian}
\end{eqnarray}
where $\state{v_n}$ represents normalized spin states determined by the MC algorithm, and $E_\theta(v_n)$ corresponds to their energy as measured by the chosen EBM ansatz. It is important to note that $\mathcal{K}_\theta$ is defined as a Hermitian operator. Using the modular Hamiltonian definition in eq.~\eqref{eq:hamiltonian}, the expectation value of $d$-th data point can be defined as
\begin{eqnarray}
    \expK = \frac{1}{N}\sum^{N}_{i=1} \stateinv{p_n}^i_d \Unitary\Kth\Unitaryd\state{p_n}^i_d \ ,\label{eq:expK}
\end{eqnarray}
where the mean expectation value has been computed by taking the mean of $N$ samples taken from the Bernoulli distribution.

\subsubsection{The objective function}

The primary objective of this endeavor is to establish a dependable representation of $\sigD$ through the acquisition of a learned density matrix, denoted as $\rthp$. In classical probabilistic learning, the optimization process revolves around minimizing the Kullback-Leibler divergence, $D_{\mathrm{KL}}$, to diminish the disparity between two probability distributions (refer to eq.~\eqref{eq:kld})~\cite{Murphy}. In this context, the goal is to approximate $\rthp \simeq \sigD$. Extending this principle to our scenario, where we work with a given Hamiltonian and temperature, the Gibbs-Delbr\"{u}ck-Moli\'{e}ve variational principle~\cite{Huber} asserts that the most suitable objective function for this process is the free energy, defined as
\begin{eqnarray}
    \mathcal{F} = E - \frac{1}{\beta}\mathcal{S}(\sigD)\ ,
\end{eqnarray}
which is bounded by the actual free energy of the system. Here, $\mathcal{S}(\sigD)$ denotes the entropy of the data (refer to eq.~\eqref{eq:vnentropy}), $\beta$ represents the inverse temperature, and $E$ signifies the expectation value of the given Hamiltonian as defined in eq.~\eqref{eq:expK}. Notably, both the Hamiltonian and the entropy are unknown prior to data analysis, posing a significant challenge. Recognizing that the free energy can also be expressed as the log-partition function, we can reformulate the entire expression as follows:
\begin{eqnarray}
    \beta E + k_\beta \log\mathcal{Z}_\theta\ \geq \mathcal{S}(\sigD)\ ,\label{eq:objective_ineq}
\end{eqnarray}
where $k_\beta$ is a constant related to the true entropy of the data and the Boltzmann constant. This inequality offers a more suitable objective function for our purpose. Since temperature and the Boltzmann constant lack physical significance in the context of data analysis, they can be employed as regularizers of the objective function, and for this study, we consider them to be equal to one.

After staring to the left-hand side of the eq.~\eqref{eq:objective_ineq}, it becomes evident that the entropy of the data essentially manifests as the negative log-probability distribution of a multivariate Gaussian distribution centered at zero:
\begin{eqnarray}
    \mathcal{L}(\theta,\phi) = \frac{1}{\mathcal{Z}_\theta^{k_\beta}}e^{-\beta \expK} \equiv \mathcal{N}\left(|\hat{U}(\phi)p_n\rangle\bigg|0, \Sigma(\theta)\right)\ , \nonumber
\end{eqnarray}
where the Hamiltonian can be interpreted as the covariance matrix ($\Sigma(\theta)$) among different features. Given that the determinant of a Hermitian matrix is equivalent to the sum of its eigenvalues, $\mathcal{Z}_\theta^{k_\beta}\equiv \det(\Sigma(\theta))^{{\rm dim}(\Sigma)}\sqrt{2\pi}$. Thus, our approach essentially models the data as a multivariate Gaussian distribution while simultaneously learning the covariance matrix through the optimization of $-\log\mathcal{L}(\theta,\phi)$. Similar analogies can be found in the context of simulating lattice field theories using flow-based algorithms~\cite{Albergo:2022qfi, Cranmer:2019kyz}.

In unsupervised learning, the neural network serves as a statistical model of the underlying data, and the objective is to minimize the negative log probability distribution. Drawing from the analogy presented earlier, it becomes evident that reformulating this problem as a thermal state effectively implies an assumption that the data can be suitably approximated by a Gaussian distribution. This insight establishes a direct connection between theoretical approaches and conventional machine learning techniques, highlighting the interplay between sophisticated modeling strategies and established methodologies in the field of statistical thermodynamics.

\subsubsection{Combining it all together}

\begin{figure*}[!ht]
    \centering
    \includegraphics[scale=0.55]{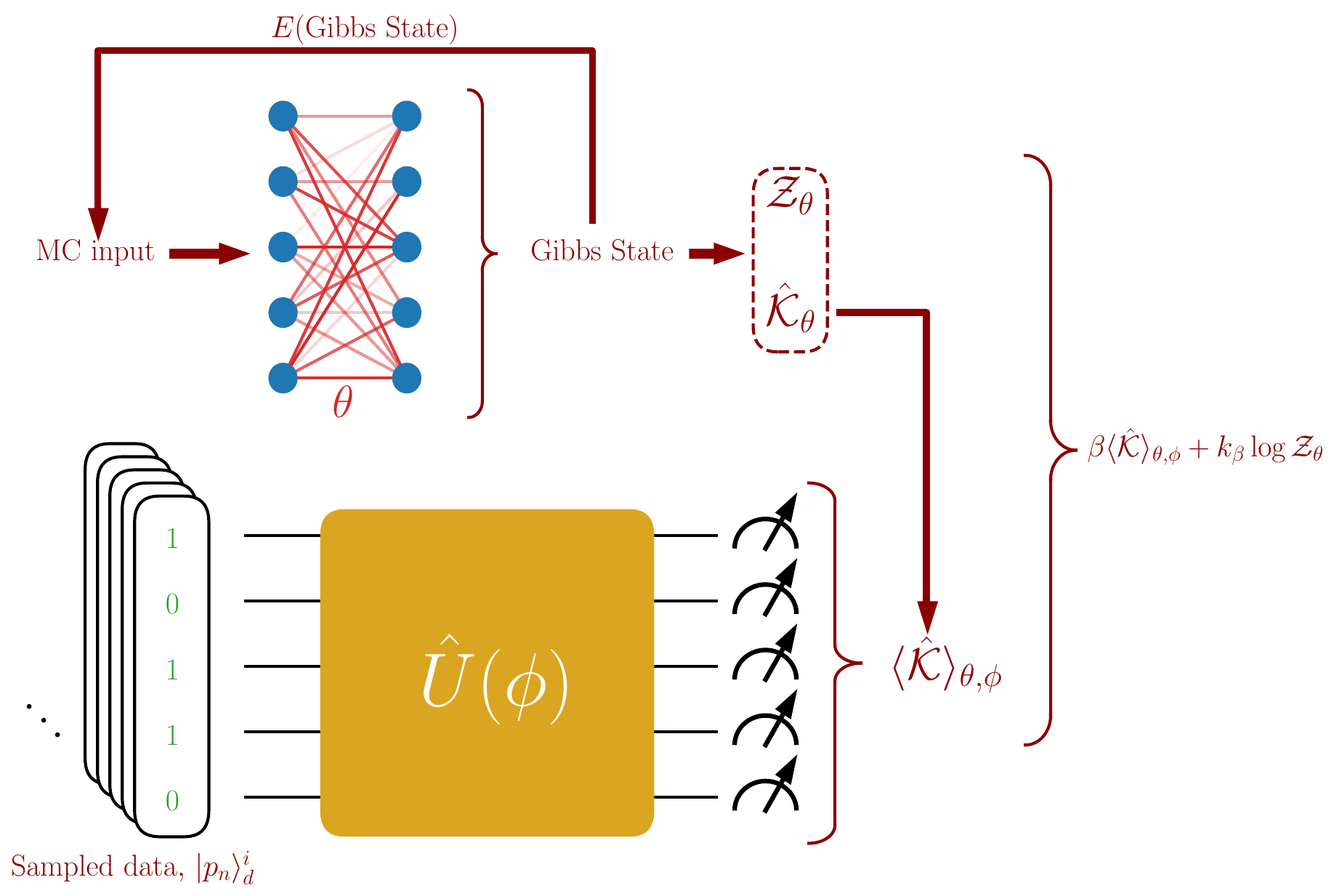}
    \caption{\it Schematic representation of the Quantum Modular Hamiltonian-based learning for data analysis. Two parts of the implementation has been represented as two parallel layers stacked on top of each other in the representation where the top layer is responsible of forming a Hamiltonian by generating a Gibbs state through a MC algorithm based on an EBM. The bottom layer is responsible to compute expectation value of the Hamiltonian for a sampled set of initial states. Finally, expectation value and the partition function is combined to form the cost function of the network.}
    \label{fig:qhbm_sketch}
\end{figure*}

Fig.~\ref{fig:qhbm_sketch} provides a schematic overview of the entire process, segmented into two primary panels. The upper panel illustrates the generation of the modular Hamiltonian, $\hat{\mathcal{K}}_\theta$, and the partition function, $\mathcal{Z}_\theta$, through the utilization of a MC algorithm. In contrast, the lower panel depicts the variational circuit, incorporating sampled data points as specified in eq.~\eqref{eq:bernoulli_sample} as input. Leveraging the modular Hamiltonian, we compute the expectation value for each data point, as detailed in eq.~\eqref{eq:expK}. Subsequently, we combine the partition function and the expectation value to formulate the loss function, as elucidated in eq.~\eqref{eq:objective_ineq}.

Once the mean loss function is computed for a batch of data points, we update the trainable parameters $\theta$ and $\phi$ using the following expressions:
\begin{eqnarray}
    \theta^\prime = \theta + \eta \frac{\partial}{\partial\theta}\log\mathcal{L}(\theta, \phi)\quad , \quad \phi^\prime = \phi - \eta\beta \frac{\partial\expK}{\partial\phi}\ , \nonumber
\end{eqnarray}
where $\eta$ denotes the learning rate. Algorithm~\ref{code:train_step} shows a single training step for a batch of data.

\begin{algorithm}[H]
 Form $\hat{\mathcal{K}}_\theta$ and $\mathcal{Z}_\theta$ with algo.~\ref{code:mc}\;
 Number of Bernoulli samples: $N_b$\;
 Probability of all features in a data point: $p$\;
 \While{$i<{\rm number\ of\ samples\ in\ a\ batch}$}{
    \While{$n<N_b$}
    {
    generate $n$-th sample from $i$-th data point\;
    compute $\langle p |^i_n \hat{\mathcal{K}}_\theta| p \rangle^i_n$ of $n$-th sample\;
    increment $n$\;
    }
    take the mean of $N_b$ expectation values\;
    increment $i$\;
  }
  take the mean of the expectation values in batch\;
  compute loss with mean batch expectation value and partition function\;
  compute gradient of the loss\;
  update $\theta$ and $\phi$\;
 \caption{\it Pseudo-code for a single training step.}\label{code:train_step}
\end{algorithm}

% As a recap, Fig.~\ref{fig:qhbm_sketch} is a schematic representation of this process. Blocks on the left contain the binary feature encoding sampled from the Bernoulli distribution as shown in eq.~\eqref{eq:bernoulli_sample}. The expectation value of this sample represents the expectation value of one data point. To find the sample's mean expectation value, each batch's expectation value, containing $N_{\rm smp}\times N$ circuits, has been averaged. The batch's entropy or negative log probability has been computed from the log-partition function and mean expectation value of the modular Hamiltonian, eq.~\eqref{eq:objective_ineq}. Finally the parameters $\theta$ and $\phi$ are updated via,

Within the concept of batched learning, optimising negative log-probability distribution has two significant computational bottlenecks. First, as mentioned above, one must estimate a proper modular Hamiltonian for the optimisation procedure. This has been done independently of the input data where the MC algorithm chooses the most probable set of spin configurations and, by doing so, automatically minimises the energy of the EBM ansatz. This requires reestimation of the modular Hamiltonian after each update. It is important to emphasise that since the MC algorithm proposes an input configuration completely independent of the input data, the optimisation procedure is the only connection between data and the Hamiltonian. The second bottleneck involves $\sigd$ estimation; since we do not have access to ``quantum data", we need to sample through the probability distribution of each data point and compute the expectation value $\expK$. 

Such an optimisation process allows the quantum circuit to learn a non-linear distribution by minimising $E_\theta(v_n)\stateinv{\sigD\hat{U}(\phi)}\state{v_n}\stateinv{v_n}\state{\sigD\hat{U}(\phi)}$. Since $\state{v_n}$ is constructed from a non-linear classical neural network, $\state{\sigD\hat{U}(\phi)}$ is being forced to approximate such non-linear behaviour to reduce the distance of the projection. 

Although it is a powerful representation of the data, learning a completely free Hamiltonian is computationally challenging simply because the Hamiltonian has to be decomposed into Pauli operators at every step of the optimisation process. It is possible to avoid the EBM if we assume a certain structure for the modular Hamiltonian $\hat{\mathcal{K}}$. For instance, a generic Hamiltonian that captures the nearest neighbour interactions can be suitable to capture near-term complexity of the data,
\begin{eqnarray}
    \hat{\mathcal{K}}_\theta = \sum_{i\in{\rm qubits}} \theta_{i,i+1}\left(\sigma^+_i\sigma^-_{i+1} + \sigma^+_{i+1}\sigma^-_i\right)\ ,\nonumber
\end{eqnarray}
where $\sigma^\pm$ are raising and lowering operators and $\theta$ are the trainable coupling strength. Since the summation captures only nearest neighbour interactions, this Hamiltonian may not be able to capture the complexity of the data, but it will simplify the optimisation process significantly. Additionally, tensor network techniques can aid in the decomposition of large Hamiltonian matrices. However, such simplifications are out of the scope of this paper, where we focus on the most generic application and see if it is indeed possible to create a useful operator through this procedure.

%%%%%%%%%%%%%%%%%%%%%%%%%%%%%%%%%%%%%%%%%%%%%%%%%%%%%%%%%%%
\section{Results}\label{sec:results}
%%%%%%%%%%%%%%%%%%%%%%%%%%%%%%%%%%%%%%%%%%%%%%%%%%%%%%%%%%%

As a case study, we used top tagging dataset~\cite{Kasieczka:2019dbj, kasieczka_gregor_2019_2603256}, which includes over a million mixed collider events for semi-leptonic top and dijet production channels at $\sqrt{s} = 14$ TeV. Events are generated and showered in \textsc{Pythia}~8~\cite{Sjostrand:2014zea}, and the detector simulation has been achieved using \textsc{Delphes}~3 package~\cite{deFavereau:2013fsa} with default ATLAS configuration card. All jets are reconstructed via \texttt{anti-kT} algorithm~\cite{Cacciari:2008gp} with $R=0.8$ within \textsc{FastJet}~\cite{Cacciari:2011ma} package. Furthermore, the central-boosted phase-space has been captured by requiring jet transverse momentum, $p_T$, to be within $[550, 650]$ GeV and absolute pseudo-rapidity to be $|\eta| < 2$.

The jets are further processed to be represented as calorimeter images, potentially captured by the hadronic calorimeter (HCAL) at an LHC experiment. Following the procedure presented in refs.~\cite{Araz:2021wp, Araz:2021un}, leading jet constituents are centred on around the jet-axis on pseudo-rapidity and azimuthal angle, $\eta-\phi$, plane within $[-1.5,1.5]$. Each image is divided into four-quadrant, and the most energetic quadrant has been moved to the top-right corner by horizontally and vertically flipping the image. Finally, all the training samples are standardised over randomly chosen 200,000 images by fitting $p_T$ within the $[0,\pi]$ range. This standardisation procedure yields calorimeter images of $37\times37$ pixels; however, since it is not possible to process this within a quantum circuit, we simplified our data by cropping 12 pixels from each axis and down-sampling the resulting image by taking the mean of four adjacent pixels. 
\begin{figure*}
    \centering
    \includegraphics[scale=0.45]{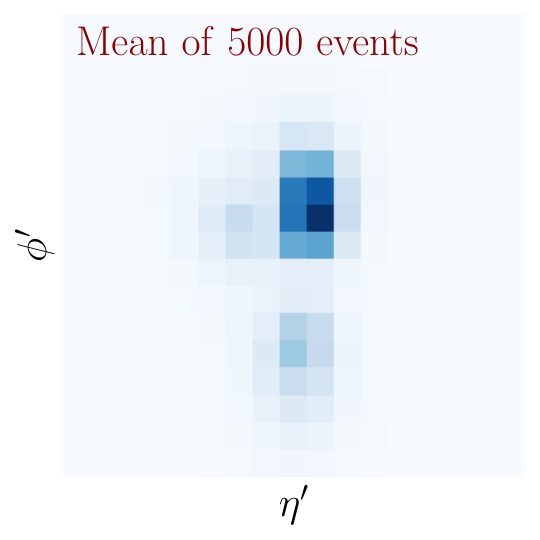}
    \includegraphics[scale=0.45]{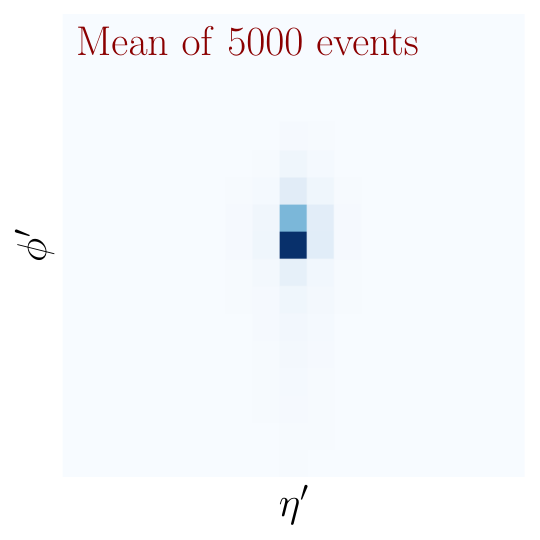}
    \includegraphics[scale=0.45]{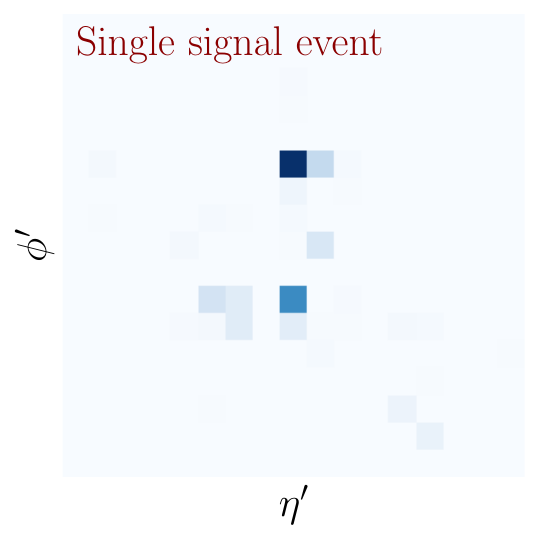}
    \includegraphics[scale=0.45]{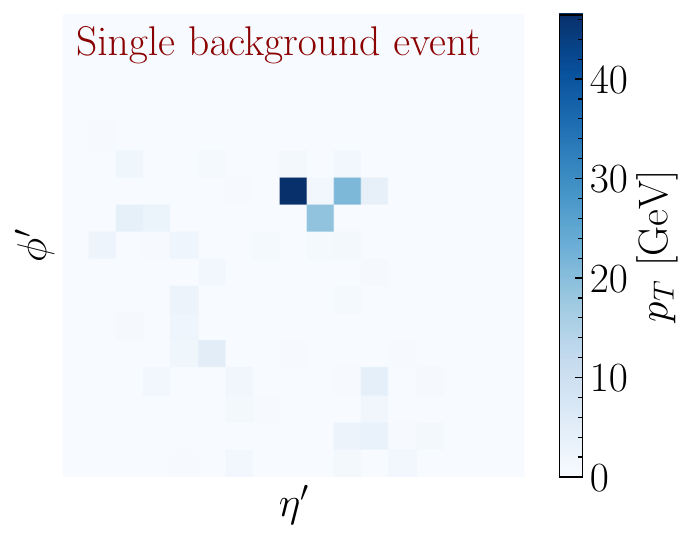}
    \caption{\it Signal and background images projected on $\eta^\prime-\phi^\prime$ frame. From left to right, images represent signal and background represented with a mean of 5000 randomly chosen events and with a single event for each sample. For this representation, ten pixels have been cropped from each axis of the images from the original $37\times37$ pixels. Colour represents the magnitude of energy deposited in each pixel.}
    \label{fig:data_image}
\end{figure*}

Two left panels of Fig.~\ref{fig:data_image} show the mean of 5000 images for signal (on the left) and background (on the right) where $\eta^\prime$ and $\phi^\prime$ are the modified pseudo-rapidity and azimuthal angle axes after standardisation. The colour represents the value of the transverse momentum in each pixel, measured on the very right panel of the figure. Notice that this is before normalizing $p_T$ distribution within $[0,\pi]$ range. The following two image captures only a single event within the signal (on the left) and background (on the right) samples. All the images are cropped to show only $27\times27$ central image to focus on the main activity. Even though the averaged images are easily differentiable, single events are usually random looking and not easily differentiable; hence various ML techniques have been developed to differentiate these samples.

Because the top decays into two light jets through a W decay and a b-jet, it creates a three-prong signature, as shown on the very left panel of Fig.~\ref{fig:data_image}. Such topological behaviour has been exploited by many analytic tagging algorithms (e.g. ref.~\cite{Plehn:2009rk}). The dijet signature, on the other hand, leaves a single prong signature on the calorimeter, as shown in the second image on the left panel of Fig.~\ref{fig:data_image}. This is due to the fact that dijet events do not contain enough energy to produce two distinct jet signatures. Such a process is crucial to investigate at the LHC because the top quark's mass can further our understanding of the Higgs mechanism and its coupling to the top quark since mass comes with a large coupling to the Higgs boson. Even with larger centre-of-mass energies, the production of top quark pairs has been improved at the LHC. These events are usually contaminated with dijet events, making it challenging to isolate top quark events. Hence it is vital to separate top events from the dijet background to improve the experiment's sensitivity to its couplings.

In the following sections, we will use only a fraction of these images by cropping and downsampling them due to computational limitations. This mainly affects the span of information since it is highly dependent on the geometrical positions of the energy deposits on each pixel. We will start with the central pixels and increase the pixel count from there, but it is important to note that the central four pixels are maximally similar throughout both samples. Hence one includes more information regarding the nature of the event once we go beyond the central four pixels.

%%%%%%%%%%%%%%%%%%%%%%%%%%%%%%%%%%%%%%%%%%%%%%%%%%%%%%%%%%%
\subsection{Generative modelling}\label{sec:generative_modeling}
%%%%%%%%%%%%%%%%%%%%%%%%%%%%%%%%%%%%%%%%%%%%%%%%%%%%%%%%%%%

As the first set of applications for QHBM, we will aim to learn the probability distribution of the pixel intensity in calorimeter images. Each standardised sample pixels has $p_T$ intensity between $[0,\pi]$. The pixel intensity can be interpreted as probability distribution if it passes through a bijective function which outputs values between $[0,1]$, such as a sigmoid function. This will allow us simultaneously interpret pixel intensities as probability distributions and convert them back to their status quo. Due to the computational cost of the quantum simulation and the optimisation methodology, we choose to perform our investigation with only the central four pixels, which retain the necessary information to differentiate between top and dijet images as presented in a previous study~\cite{Araz:2022haf}.

The modular Hamiltonian has been determined via Restricted Boltzmann Machine (RBM), where details have been presented in App.~\ref{app:rbm}. We reestimate the modular Hamiltonian for each batch training by collecting a set of spin states via the MC algorithm presented above. The initial state for each training has been set to $\state{\uparrow\cdots\uparrow}$; each following MC algorithm has been initiated by the last state determined in the previous MC run. For each execution MC algorithm ran for 100 steps to converge on a stable Gibbs state without collecting any; the number of collected states is analysed case by case below. Note that these states are entirely independent of the input data; hence MC algorithm independently minimises the energy of the RBM by choosing the most probable set of states.

The expectation value for each image has been estimated via eq.~\eqref{eq:expK}. Since we are employing batched learning, the expectation value of the batch has been computed by taking the mean of each expectation estimation in the batch. Finally, the variational parameters of the network have been updated with respect to the mean objective function,
\begin{eqnarray}
    \arg\min_{\theta,\phi} \frac{1}{N_{\rm batch}} \sum^{N_{\rm batch}}_i-\log\mathcal{L}(\theta, \phi|\hat\sigma_i) \ .\nonumber
\end{eqnarray}
Notice that the mean only entitles the expectation value of the modular Hamiltonian. We divided our study into different benchmarks to study the effects of $\sigD$ and $\Kth$ estimations. \textsc{PennyLane} package~\cite{bergholm2020pennylane} has been employed for quantum circuit simulation, the RBM and optimisation have been held within \textsc{TensorFlow}~\cite{DBLP:journals/corr/AbadiBCCDDDGIIK16, tensorflow2015-whitepaper}, and \textsc{TensorFlow-Probability}~\cite{https://doi.org/10.48550/arxiv.1711.10604} packages. Our implementation can be found in this \href{https://gitlab.com/jackaraz/qhbm}{GitLab repository}\footnote{\href{https://gitlab.com/jackaraz/qhbm}{https://gitlab.com/jackaraz/qhbm}.}. All the benchmarks are trained with 1000 training samples, and overtraining has been monitored with the same number of validation events\footnote{It is essential to note that we did not observe any significant improvement in generalisation for more extensive training sets; hence, due to the computational cost, we limited the analysis to 1000 event.}. \texttt{Adam} optimisation algorithm~\cite{Kingma2014AdamAM} has been employed with $10^{-2}$ initial learning rate, where the learning rate has been reduced to its half if validation loss has not been improved for over 25 epochs. Each benchmark has been trained for 100 epochs, and training has been terminated if the validation loss hasn't been improved for over 50 epochs.
\begin{figure}
    \centering
    \includegraphics[scale=0.45]{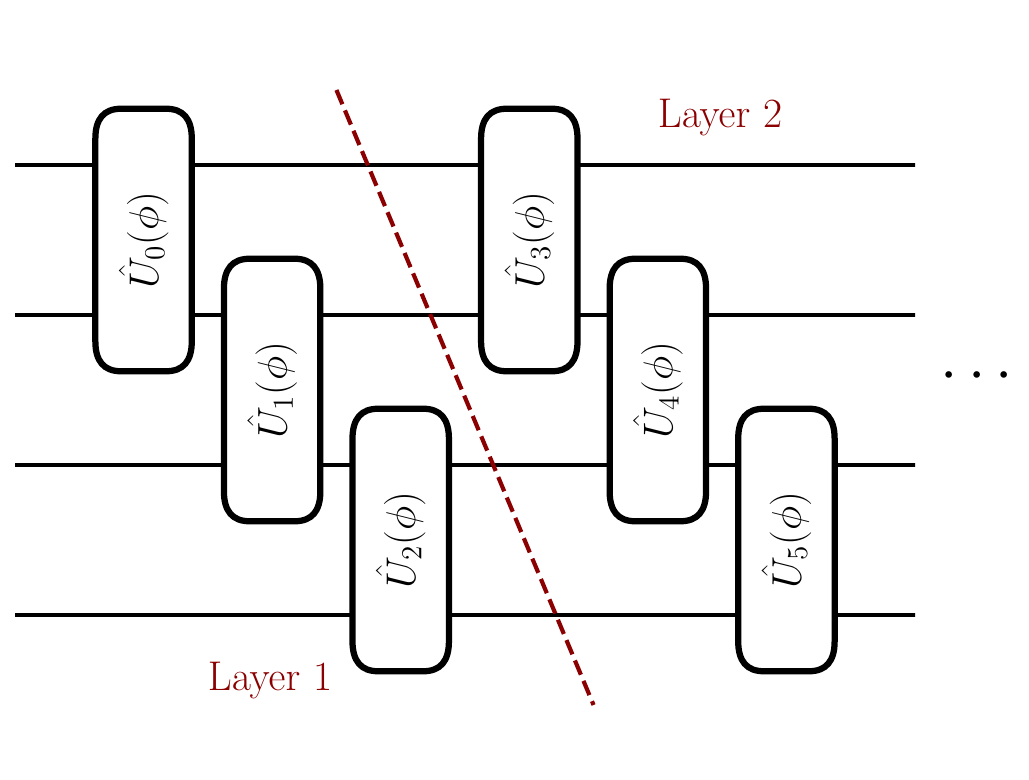}
    \includegraphics[scale=0.45]{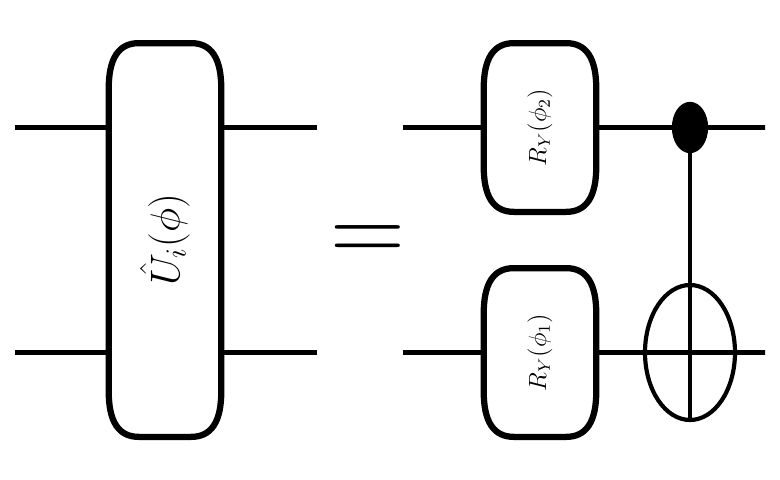}
    \caption{\it Upper panel shows a schematic diagram of the MPS variational circuit ansatz, and the bottom panel shows the structure of each $\hat{U}(\phi)$ gate that has been used.}
    \label{fig:mps_ansatz}
\end{figure}

For the quantum ansatz, we used Matrix Product state (MPS) structure~\cite{Huggins_2019} where two-qubit operators have been applied to each adjacent qubit in a staircase-like architecture which is depicted in Fig.~\ref{fig:mps_ansatz}. We will refer to each of these constructions from the first qubit to the last as a layer. Each two-qubit operator, $\hat{U}_i(\phi)$, includes two rotation gates around the Pauli-Y axis for each input qubit with an independent variational rotation angle followed by a CNOT gate. For each benchmark, we used three layers. Note that the algorithm has also been tested with different architectures such as simplified two-design~\cite{Cerezo:2021aa} and strongly entangling layers~\cite{Schuld2020}, which has been observed to improve the results.
\begin{figure*}[!ht]
    \centering
    \includegraphics[scale=0.56]{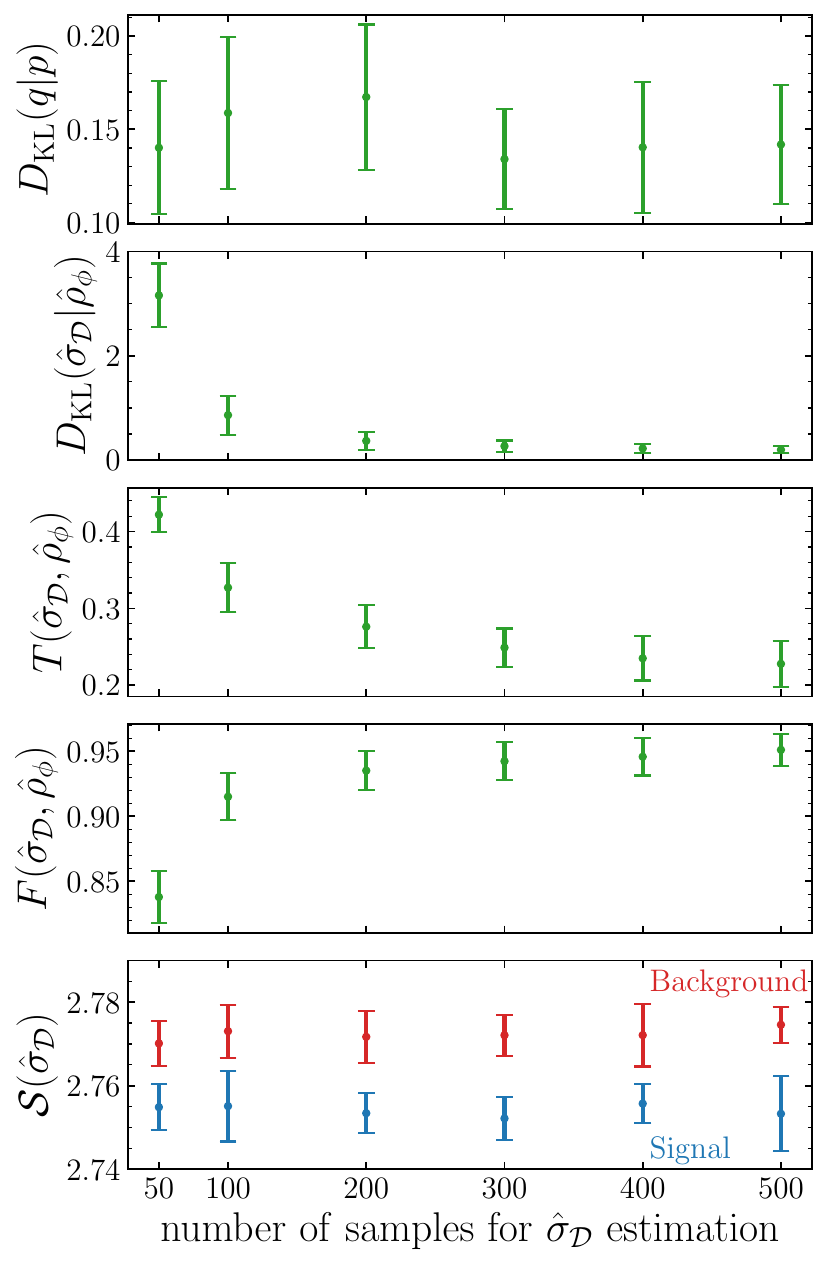}\quad \quad \quad
    \includegraphics[scale=0.56]{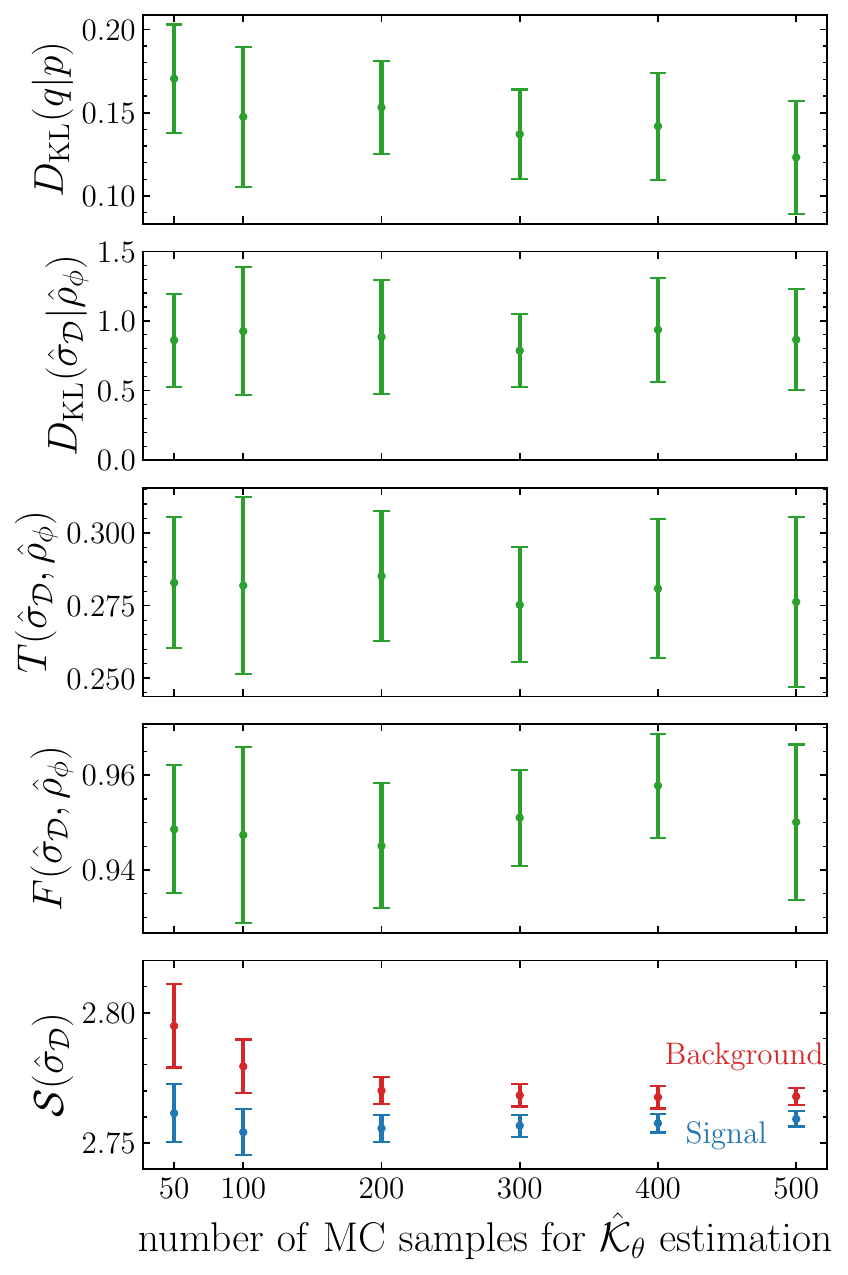}
    \caption{\it Test metrics presented for various networks where the left panel shows the network trained with a different number of samples for density matrix estimation and the right panel shows the network trained with a different number of MC samples for $\hat{\mathcal{K}}_\theta$ estimation. The top panels of both sides show the KL divergence between input and output samples, followed by KL divergence, trace distance and fidelity between the truth level density matrix and the network's density matrix estimation. The bottom panel shows the network's estimation for von Neumann entropy, where red and blue represent background and signal samples. The results on the left panel are prepared using 200 MC samples for the estimation of the Hamiltonian. Similarly, the right panel uses 5000 samples for $\hat{\sigma}_\mathcal{D}$ estimation. Each result has been presented with one standard deviation, estimated by dividing 10,000 test samples into batches of 25 events.}
    \label{fig:test_metrics}
\end{figure*}

Fig.~\ref{fig:test_metrics} shows the test metrics for each benchmark where each point has been tested with 10,000 mixed test events and presented with one standard deviation, estimated by dividing the test sample into batches of 25. The left panel shows the benchmarks for $\sigD$ estimation where $N^{MC}$ for $\Kth$ estimation have been set to 200. On the other hand, the right panel shows the benchmarks for $\Kth$ estimation where $N^{\rm smp}$ for $\sigD$ estimation have been set to 5000. It is also important to note that the samples to estimate $\sigD$ for the right panel are generated before the optimisation process to speed up the application; however, for the left panel, each sample produced during the training effectively allowed those benchmarks to see different samples in each iteration.
Each panel is divided into five sub-panels, where the top two panels present the Kullback-Lieber distance between input images ($q$) and the sampled output images ($p$) and batched mixed state of the data ($\sigD$) and estimated mixed state ($\hat\rho_\phi$)\footnote{Notice the difference in notation here, for the variational thermal state we used $\rthp$ during the training since the density matrix at this stage, has been influenced both by the EBM and $\Unitary$. However, during the testing, modular Hamiltonian haven't been used; thus, the variational thermal state has only been influenced by $\Unitary$.}. In the following panels, we present trace distance and fidelity between the thermal state of the data and the estimated thermal state. Finally, we plotted the estimation of the von Neumann entropy separately for the signal (blue) and the background (red)\footnote{For gray scale print; signal is represented with dark gray where background is light gray throughout the paper.}. Note that for the thermal state of the data all images in the input has been assigned the same weight i.e. $\alpha_i$ in eq.~\eqref{eq:sigma_n}. For the details about these metrics, we refer the reader to App.~\ref{app:metrics}.

During our tests, we observed that the performance of the generative model is mainly based on the wellness of the estimation of the mixed state of the data, where for larger samples, we observed improved fidelity and trace distance (see the left panel of Fig.~\ref{fig:test_metrics}). Furthermore, we observed that the wellness of the estimation also improves the Kullback-Lieber distance between the input and estimated states and exponentially reduces this metric's uncertainty. Notice that $\mathcal{S}(\sigD)$ has been presented separately for signal and background. Although we haven't seen any significant difference in signal or background for any other metric, the entropy for different sample sets has been clearly separated. Note that all benchmarks are trained with mixed data, and neither has been exposed to the information regarding the data type. 

On the right panel, we present the effect of the $\Kth$ estimation on the same test metrics. Although we haven't observed any significant improvement in fidelity, trace distance and Kullback-Lieber distance (except a minor refinement in $D_{\rm KL}(q|p)$), we observed that wellness of $\Kth$ estimation improves the entropy estimation of the data and reduces the uncertainty. Hence the bottom right panel of Fig.~\ref{fig:test_metrics} indicates that for good enough $\Kth$ and $\sigD$ estimation, signal and background samples will produce unique entropy values. Thus this information can also be used to identify the nature of the data. However, $\mathcal{S}(\sigD)$ has not been observed to be a powerful discriminator. We computed the receiver operating characteristic curve to quantify the difference between signal and background, and the highest area under the curve value we observed was around 0.7.

Note that we haven't discussed the advantage of learning an operator for the data. In the following section, we will discuss a possible usage of the modular Hamiltonian in the context of anomaly detection.

%%%%%%%%%%%%%%%%%%%%%%%%%%%%%%%%%%%%%%%%%%%%%%%%%%%%%%%%%%%
\subsection{Anomaly detection}\label{sec:anomaly}
%%%%%%%%%%%%%%%%%%%%%%%%%%%%%%%%%%%%%%%%%%%%%%%%%%%%%%%%%%%

Anomaly detection is a methodology in which the network ansatz learns the structure of the known data and tries to detect the difference in new data, if any. For this purpose, we have used two test cases. For the first case, we used six qubits where in addition to the central four pixels, we added the top two pixels into the collection. For the second case, we also included the bottom two pixels to test the algorithm for eight qubit scenario.

We are using the same procedure outlined in sec.~\ref{sec:generative_modeling}, trained both scenarios using background-only samples for 100 epochs and 1000 events where $\sigD$ is estimated by 5000 samples before the training. The only difference between the two test cases is that we used 500 MC samples for six qubit scenario and 1000 for eight qubit scenario. The difference is due to the size of the latent space, where we observed that a larger latent space requires more MC samples to estimate $\Kth$ for the stability of the result, which we will discuss later in this section.
\begin{figure*}[!ht]
    \centering
    \includegraphics[scale=0.51]{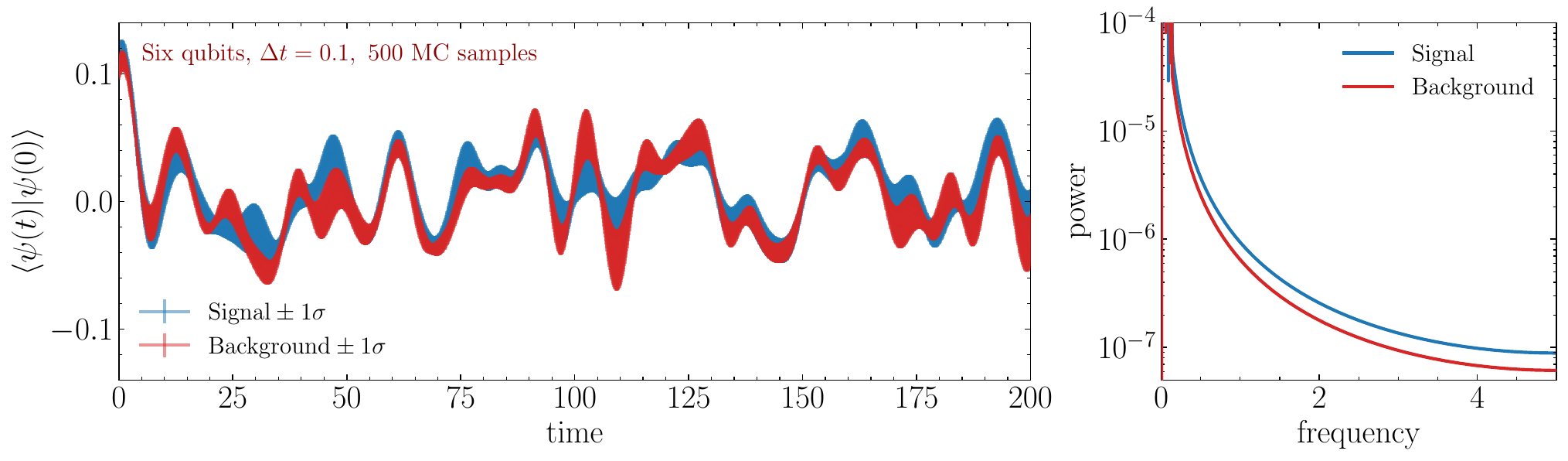}
    \includegraphics[scale=0.51]{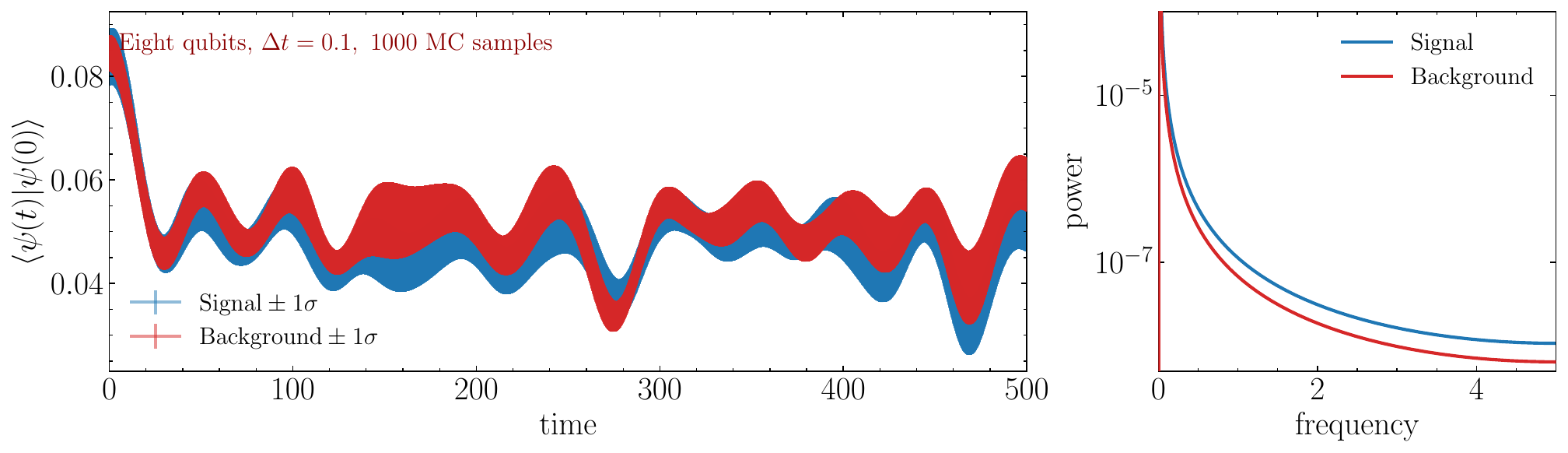}
    \caption{\it Time evolution of the modular Hamiltonian for six qubits (top panel) and eight qubits (bottom panel) scenario. The left panel shows the fidelity distribution, and the right panel shows the power spectrum of the FFT of the distributions. The signal and the background are represented with blue and red colours in each panel.}
    \label{fig:time_evo}
\end{figure*}

The network results have been tested with 10,000 background-only test samples. For the six-qubit scenario, we observed fidelity of 0.81 and a trace distance of 0.3, whereas, for eight qubit scenario, we observed 0.79 and 0.3, respectively. 

Although von Neuman entropy, as shown in sec.~\ref{sec:generative_modeling}, can lead to a significant observable to differentiate two types of samples, we propose a new observable based on the modular Hamiltonian.  We will analyse two different cases; for the first, we will look into the effect of time evolution. We will define the time evolution operator of a modular Hamiltonian as 
\begin{eqnarray}
    e^{-iT\Kth} \simeq \prod^{N}e^{-i\Delta t \Kth} \equiv \mathcal{T}_N\ , \nonumber
\end{eqnarray}
where $T = N\Delta t $. For small $\Delta t$, this operator can be applied on the quantum circuit under the Trotter-Suzuki approximation. Using this relation, one can compute the fidelity of the time-evolved quantum state as
\begin{eqnarray}
    {\rm Fidelity} = \langle \psi(t) | \psi(0) \rangle\ , \label{eq:time_evo_fid}
\end{eqnarray}
where $| \psi(0) \rangle = \hat{U}(\phi) |p_n\rangle^i_d$ and $|\psi(t)\rangle = \mathcal{T}_N | \psi(0) \rangle$. For the second case, since its computationally less costly, we will analyse the expectation value of the without time evolution. 

\begin{figure*}[!ht]
    \centering
    \includegraphics[scale=0.55]{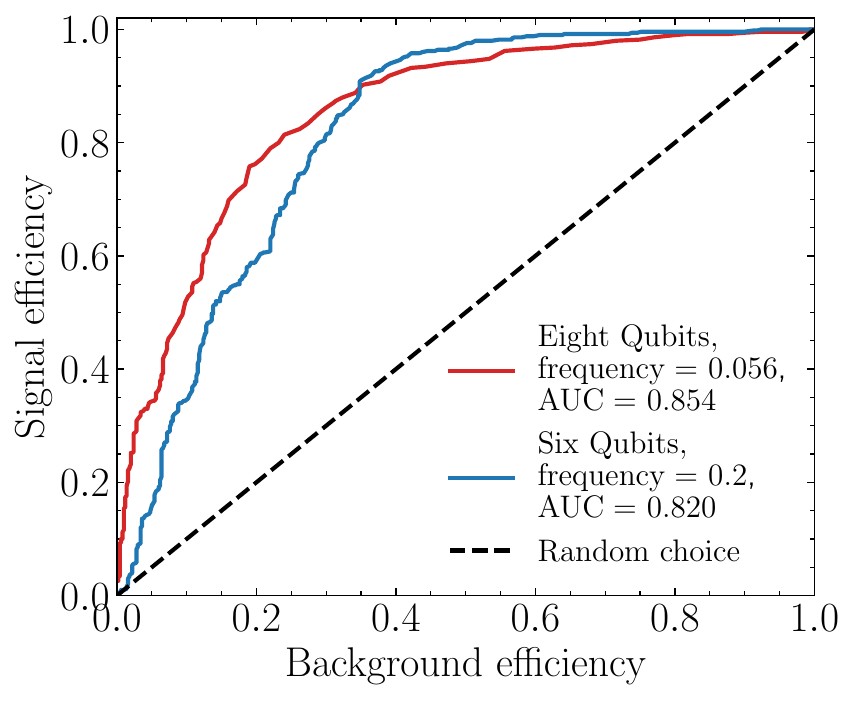}\quad \quad \quad
    \includegraphics[scale=0.55]{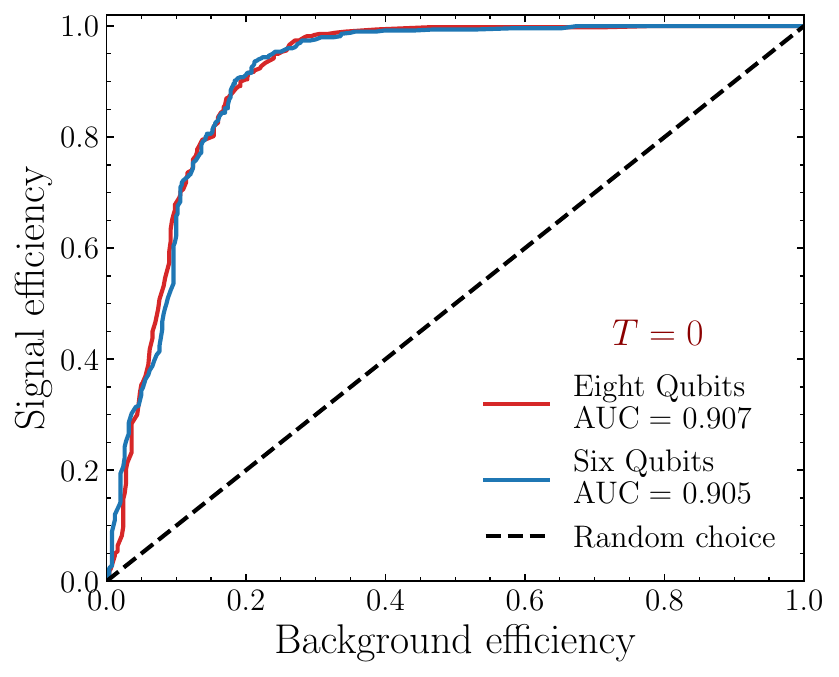}
    \caption{\it ROC curve computed for two type measures been used in this study. The left panel shows the results produced for the time evolution sequence of the learned Hamiltonian, and the right panel shows the results for the expectation value at $T=0$. The red and blue colours represent eight and six-qubit scenarios, and the dashed black curve shows the random choice.}
    \label{fig:roc_anomaly}
\end{figure*}

 We computed the time evolution up to $T\leq 500$ for estimated $\Kth$ in each scenario with $\Delta t = 0.1$ time steps. The left panel of Fig.~\ref{fig:time_evo} shows the fidelity, eq.~\eqref{eq:time_evo_fid}, concerning each time step for signal (blue) and background (red) samples where the six-qubit scenario is presented in the upper panel and the eight-qubit scenario in the lower panel. The thickness of each curve shows one standard deviation for the entire test sample\footnote{Note that the test sample has been limited due to the high computational cost.}. Note that for the sake of visibility, plot referring to six qubit scenario has been limited to  $T\leq200$ whilst the computation has been done for $T\leq500$. In order to devise a quantitative measure, we computed the power--frequency curve from the fast Fourier transform of the time evolution sequence (see eq.~\eqref{eq:power}). For the mean time-evolution sequence, we present power--frequency distribution on the right panels of Fig.~\ref{fig:time_evo} for each respective time-evolution result. Although we haven't observed any significant difference in low-frequency regions, the power of both curves becomes significantly different for high-frequency regions. It is essential to note that the power--frequency curve becomes identical once the network is trained with mixed signal and background samples. Additionally, for the four qubit scenario, the differentiability has been observed to be significantly low. We compute the receiver operating characteristic (ROC) curve concerning the power distribution for a frequency threshold to quantify the ability to differentiate between two samples via the time evolution sequence. The true (false) positive rate, {\it i.e.} signal (background) efficiency, has been computed by counting the number of events in binned power distribution between its maximum and minimum values for a given frequency. The left panel of fig.~\ref{fig:roc_anomaly} shows the ROC curve and corresponding area under the curve (AUC) values for six (blue) and eight-qubit (red) scenarios. The dashed black line shows the random choice where the classification quality improves as the curves move further away from this line towards the upper left corner of the plot. The best minimum frequency value has been chosen for both distributions; hence we did not observe any improvement in the AUC value for larger frequencies. We observe that the eight-qubit scenario reaches saturation at a frequency of $0.056$ with a $0.85$ AUC value. In contrast, the six-qubit scenario requires a frequency of $0.2$ to reach saturation at $0.82$ AUC value.

For the second, less costly method, we compared the expectation value for signal and background without any time evolution step, $T=0$. The right panel of Fig.~\ref{fig:roc_anomaly} shows the ROC curve computed for 200 different thresholds chosen between maximum and minimum expectation values. We tested the results for a $10,000$-event signal and background test sample where, as before, the red and blue curves show the results for eight and six-qubit scenarios, respectively. Even at the initial time step, we observe that AUC values for both cases are above 0.9. 

Utilising the time evolution sequence, we observe up to 3\% difference between six and eight qubit scenarios, reducing the required frequency by 72\%. Notice that we are barely able to achieve 50\% using a four-qubit scenario with the largest frequency that we compute; thus, adding new information significantly affects the ability to differentiate two sequences. Using only the information from the expectation value provides significantly better differentiability, whereas, in the six (eight) qubit scenario, we observed 9\% (6\%) improvement in AUC values.

As mentioned before, the stability of the results relies on sufficient MC samples for $\Kth$ estimation. Due to the probabilistic nature of EBM, the computation of $\Kth$ leads to a slightly different modular Hamiltonian; hence the stability depends on increasing the number of samples; in other words, it depends on reaching a stable Gibbs state. For a lower number of MC samples, we observed a more significant standard deviation in each sample and lower differentiability between two sets of samples where the AUC value was significantly lower. 

\begin{figure*}[!ht]
    \centering
    \includegraphics[scale=0.6]{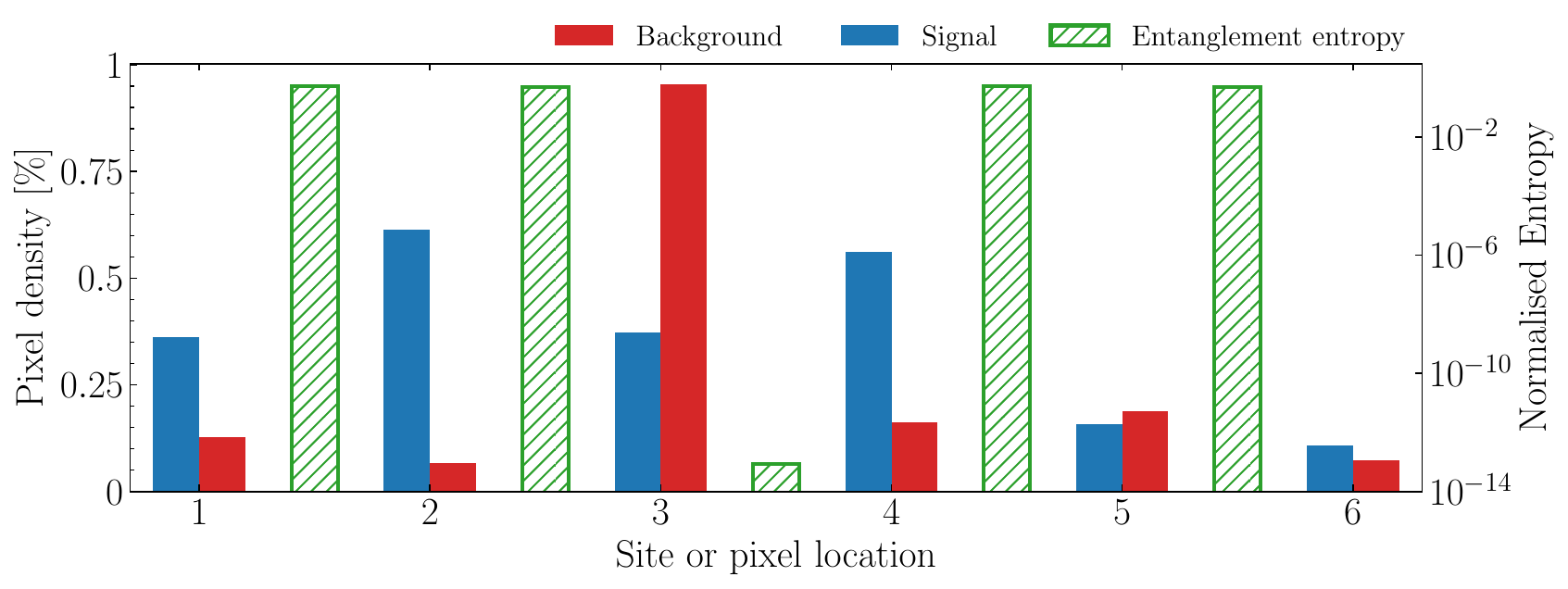}
    \caption{\it Solid bars show the pixel density of each site over the entire test data where the value is bound to the left y-axis. Hashed bars show the relative entropy, computed from the lowest eigenvector of the learned Hamiltonian, between each site and the value, which is bound to the right y-axis. The x-axis shows the site or pixel location.}\label{fig:pix_ent}
\end{figure*}

This exercise shows that the data from different sources can be interpreted as distinct quantum states; hence their corresponding Hamiltonian will produce different results when it acts on different states produced by these data samples. Since the Hamiltonian should be able to capture the entropic probability density of the given data, we investigate von Neumann entropy between each site at the ground state of the learned Hamiltonian. The reason for using von Neumann entropy is that it captures the information flow between reduced density matrices, and the change in the entropy value indicates statistically viable information for the optimisation process. This measure has also been utilised in ref.~\cite{Araz:2021un} to compress the feature space with an MPS ansatz. Von Neumann entropy has been computed by first finding the lowest eigenvector of the six-qubit learned Hamiltonian via direct diagonalisation.
Furthermore, we constructed the reduced density matrix between two sites corresponding to each pixel. Fig.~\ref{fig:pix_ent} shows the pixel density averaged over the test set for signal (blue) and background (red) bars captured by the left y-axis. This shows which pixels are statistically more active. The green hashed bar shows the relative entropy between two sites where the right y-axis has captured the value. The x-axis shows the location of each pixel on the circuit, and the green bars are placed in between each pixel location. We observe that the entropy values remain high between the low-density pixels. However, we observe exponentially low entropy values between pixels three and four, where pixel three has the highest density among all background pixels. It is essential to emphasise here that the learned Hamiltonian does not have any access to the input data, and it is constructed by generating a Gibbs state through an MC algorithm. Hence the only link between the data and the Hamiltonian is the optimisation algorithm which enables the Hamiltonian to capture the statistical distribution of the input.

%%%%%%%%%%%%%%%%%%%%%%%%%%%%%%%%%%%%%%%%%%%%%%%%%%%%%%%%%%%
\section{Discussion \& Conclusion}\label{sec:conclusion}
%%%%%%%%%%%%%%%%%%%%%%%%%%%%%%%%%%%%%%%%%%%%%%%%%%%%%%%%%%%

Quantum Hamiltonian-Based Models are a group of ansatz that attempts to approximate the probability distribution of the data by representing it as a thermal state of a learned Hamiltonian. In this context, the computationally intensive Hamiltonian learning has been mitigated to a classical network, and a variational quantum circuit has been optimised with respect to the expectation value of the learned Hamiltonian. This method is a generalisation over the Variational Quantum Thermaliser technique, where one generates the thermal state of a given Hamiltonian at a target temperature. However, using a specific Hamiltonian for an ML application will be highly constrained since it is not always possible to a priori know the correlation structure of a given data. Hence, it has been learned during the optimisation process by utilising a classical Energy-Based Model. This enables us to create a unique Hamiltonian for the data, which can then be used to scrutinise the properties of the data further. Thus this study demonstrates that the methods developed for quantum simulations are flexible and reusable for ML applications; hence shows the strong link between theoretical approaches and statistical ML techniques. This can lead to a more interpretable and intuitive ansatz by virtue of our knowledge of quantum theory, and this study aims to take a step further to achieve a fully theory-driven ML technique.

In this study, we demonstrate the usage of QHBM for generative learning and anomaly detection for LHC data. We showed that the calorimeter images could be embedded into quantum circuits as a mixed state, and a variational thermal state of a learned Hamiltonian can represent their probability distribution. As a by-product of the optimisation process, the objective function converges to the entropy of the data, which has been observed to produce unique values for different types of data. Hence, this information can be further used to identify the generated data samples.

It is essential to ask if it is possible to use the learned Hamiltonian to understand the data structure further. We have presented two possible use cases of the learned Hamiltonian for anomaly detection. For the first case, we analysed the expectation value of the time evolution sequence for the learned Hamiltonian. We showed that by converting the sequence to the frequency domain, one could observe significantly different curves for two types of samples by computing the power distribution for the fast Fourier-transformed sequence. Secondly, we showed that even the expectation value of the learned Hamiltonian is significantly different for different data types, which we quantify by analysing the difference at various thresholds. 

Our findings signify a fundamental property of the quantum many-body Hamiltonian. Once learned, the given Hamiltonian represents the dynamical properties of a specific quantum state. Since signal and background samples form significantly different state representations, a Hamiltonian designed for one type of sample reacts differently to a different system; since these systems have distinct dynamical properties. Hence we show that it is possible to treat a given data sample as a quantum many-body system, and by using theory-driven optimisation techniques, one can learn this system's Hamiltonian to be used to understand its properties. Although we only show two possible use cases for generative modelling and anomaly detection, we hope that such approaches can be taken to devise more interpretable ML applications and build dedicated optimisation algorithms that can utilise the system's physical properties. 

Although the usage of the quantum theory comes with significant advantages, it is essential to admit that this method comes with undeniable computational costs and limitations. The elephant in the room is the ability to execute these quantum circuits within a quantum device. Although we used a relatively small number of qubits, since generating the mixed state of each data point within the circuit requires many executions, we could not reproduce these results within a current quantum device. However, this can be improved by storing the inputed mixed states within a quantum memory device, which alleviates the need to regenerate such a computationally expensive process. As presented in the anomaly detection example, for this particular dataset, the geometrical position of the active pixels is crucial to characterise the dataset. Hence increasing the number of qubits will allow more information, and our experiments indicated that it would allow for the simulation of lower time steps for discrimination. Increasing the number of qubits also requires the implementation of extensive correlations between features. An MPS ansatz was suitable enough since our experiments were implemented with a few features. Still, we observed significant gains when more complex circuit architectures were implemented, which will be increasingly important with the implementation of larger feature spaces and makes the classical computation of the circuit increasingly challenging.

A further obstacle to the method is the completely free modular Hamiltonian which significantly affects the algorithm's scalability. With the increasing qubit size, the modular Hamiltonian grows exponentially via $2^{N_q}\times2^{N_q}$, which makes it quite challenging to scale the algorithm for larger systems. As we discussed before, this can be avoided by imposing certain assumptions on the modular Hamiltonian to limit its shape.

%%%%%%%%%%%%%%%%%%%%%%%%%%%%%%%%%%%%%%%%%%%%%%%%%%%%%%%%%%%
\section*{Acknowledgement}

We thank Ongun Ar{\i}sev and Soner Albayrak for delightful discussions. JYA acknowledges the hospitality of the Galileo Galilei Institute.

%%%%%%%%%%%%%%%%%%%%%%%%%%%%%%%%%%%%%%%%%%%%%%%%%%%%%%%%%%%

\bibliography{bibliography}

\appendix
\section{Restricted Boltzmann Machine}\label{app:rbm}

The restricted Boltzmann Machine (RBM) is a generative network which learns the joint probability distribution that maximizes the log-likelihood function~\cite{10.1162/089976602760128018, 10.1162/neco.2006.18.7.1527, doi:10.1126/science.1127647}. Compared to the generic Boltzmann Machines, RBM is formed as an undirected, asymmetrical bipartite graph with two layers, {\it i.e.} visible and hidden where all visible nodes are connected to all hidden nodes. The energy of the RBM is defined as
\begin{eqnarray}
    E(v,h) &=& -\sum_i \mathcal{B}^{vis}_i v_i - \sum_j \mathcal{B}^{hid}_j h_j \nonumber\\ &-& \sum_{i,j} v_i h_j \mathcal{W}_{ij}\ , \label{eq:boltzmann_energy}
\end{eqnarray}
where $\mathcal{B}^{vis}$ and $\mathcal{B}^{hid}$ stands for visible and hidden biases, $\mathcal{W}_{ij}$ is the weight matrix between visible and hidden state. Finally $h$ and $v$ stands for hidden (or latent) and visible (or input) states. Each configuration of the visible state is associated with a scalar energy measure, eq.~\eqref{eq:boltzmann_energy}, which measures the compatibility of a given visible state where high energy stands for low compatibility. The goal of an energy-based model (EBM) is to minimize the predefined energy function.

A hidden state is constructed with respect to the given visible state where the probability of the hidden state being one is given as
\begin{eqnarray}
    p(h|v; \theta) = \sigma(v\mathcal{W} + \mathcal{B}^{hid})\ , \nonumber
\end{eqnarray}
where $\theta$ stands for the collection of the trainable parameters presented in $\mathcal{W},\ \mathcal{B}^{hid}$ and $\mathcal{B}^{vis}$ and $\sigma$ stands for the sigmoid function. In order to construct hidden states, $h$, one sample from Bernoulli distribution with probability $p(h|v; \theta)$. Similarly, the reconstruction probability of the visible state is given by
\begin{eqnarray}
    p(v|h; \theta) = \sigma(h\mathcal{W}^T + \mathcal{B}^{vis})\ . \nonumber
\end{eqnarray}

\section{Metrics}\label{app:metrics}
The fidelity of two matrices is given by
\begin{eqnarray}
    F(\sigma, \rho) = \left\{ \tr{\sqrt{\sqrt{\sigma}\rho\sqrt{\sigma}}}\right\}^2\ . \label{eq:fidelity}
\end{eqnarray}
The trace distance between two matrices is defined as
\begin{eqnarray}
    T(\sigma, \rho) = \frac{1}{2} \tr{\sqrt{(\sigma - \rho)^\dagger (\sigma - \rho)}}\ . \label{eq:trace_distance}
\end{eqnarray}
The von Neumann entropy of a matrix is given as,
\begin{eqnarray}
    \mathcal{S}(\sigma) = \tr{\sigma \log(\sigma)}\ . \label{eq:vnentropy}    
\end{eqnarray}
The Kullback--Leiber divergence between two probability distributions (or two density matrices) is defined as,
\begin{eqnarray}
    D_{KL}(p|q) = p \log(p) - p\log(q)\ . \label{eq:kld}
\end{eqnarray}
The power of the fast Fourier transform is defined as~\cite{Cooley1965AnAF}
\begin{eqnarray}
    {\rm power}(\lambda) = \Re\left[ 2\frac{\Delta t^2}{T} ||{\rm FFT}(\lambda - \bar\lambda)||^2 \right] \ , \label{eq:power}
\end{eqnarray}
where $\lambda$ is the signal in question, $\bar\lambda$ stands for the mean of the signal, FFT stands for the fast Fourier transform, and $T = N\Delta t$ where $N$ stands for the number of time iterations with $\Delta t$ separation.

\end{document}